\documentclass[prl, showpacs, superscriptadress, twocolumn, nofootinbib]{revtex4-2}
\usepackage{graphicx}
\usepackage{amsmath}
\usepackage[left=3cm, right=3cm, top=2.5cm, bottom=2.5cm]{geometry}
\usepackage{verbatim}
\usepackage[colorlinks=True,linkcolor=red,citecolor=blue,urlcolor=blue]{hyperref}
\usepackage{tikz}

\usepackage{xspace}
\newcommand*{\algo}{\texttt{AriadneVR}\xspace}

\begin{document}

\title{Virtual Reality for Understanding Artificial-Intelligence-driven Scientific Discovery with an Application in Quantum Optics
}
\author{Philipp Schmidt}
\email{ph.s.schmidt@protonmail.com}
\author{Sören Arlt}
\email{soeren.arlt@mpl.mpg.de}
\affiliation{Max Planck Institute for the Science of Light, Erlangen, Germany}
\author{Carlos Ruiz-Gonzalez}
\affiliation{Max Planck Institute for the Science of Light, Erlangen, Germany}
\author{Xuemei Gu}
\affiliation{Max Planck Institute for the Science of Light, Erlangen, Germany}
\author{Carla Rodríguez}
\affiliation{Max Planck Institute for the Science of Light, Erlangen, Germany}
\author{Mario Krenn}
\email{mario.krenn@mpl.mpg.de}
\affiliation{Max Planck Institute for the Science of Light, Erlangen, Germany}

\begin{abstract}
     Generative Artificial Intelligence (AI) models can propose solutions to scientific problems beyond human capability. To truly make conceptual contributions, researchers need to be capable of understanding the AI-generated structures and extracting the underlying concepts and ideas. 
    When algorithms provide little explanatory reasoning alongside the output, scientists
    have to reverse-engineer the fundamental insights behind proposals based solely on examples.
    This task can be challenging as the output is often highly complex and thus not immediately accessible to humans.
    In this work we show how transferring part of the analysis process into an immersive Virtual Reality (VR) environment can assist researchers in developing an understanding of AI-generated solutions.
    We demonstrate the usefulness of VR in finding interpretable configurations of abstract graphs, representing Quantum Optics experiments.
    Thereby, we can manually discover new generalizations of AI-discoveries as well as new understanding in experimental quantum optics.
    Furthermore, it allows us to customize the search space in an informed way - as a human-in-the-loop - to achieve significantly faster subsequent discovery iterations.
    As concrete examples, with this technology, we discover a new resource-efficient 3-dimensional entanglement swapping scheme, as well as a 3-dimensional 4-particle Greenberger-Horne-Zeilinger-state analyzer.
    Our results show the potential of VR for increasing a human researcher's ability to derive knowledge from graph-based generative AI that, which is a common abstract data representation used in diverse fields of science.
\end{abstract}
\maketitle
\subsection*{Introduction}

    \begin{figure*}
        \centering
        \includegraphics[width=1\textwidth]{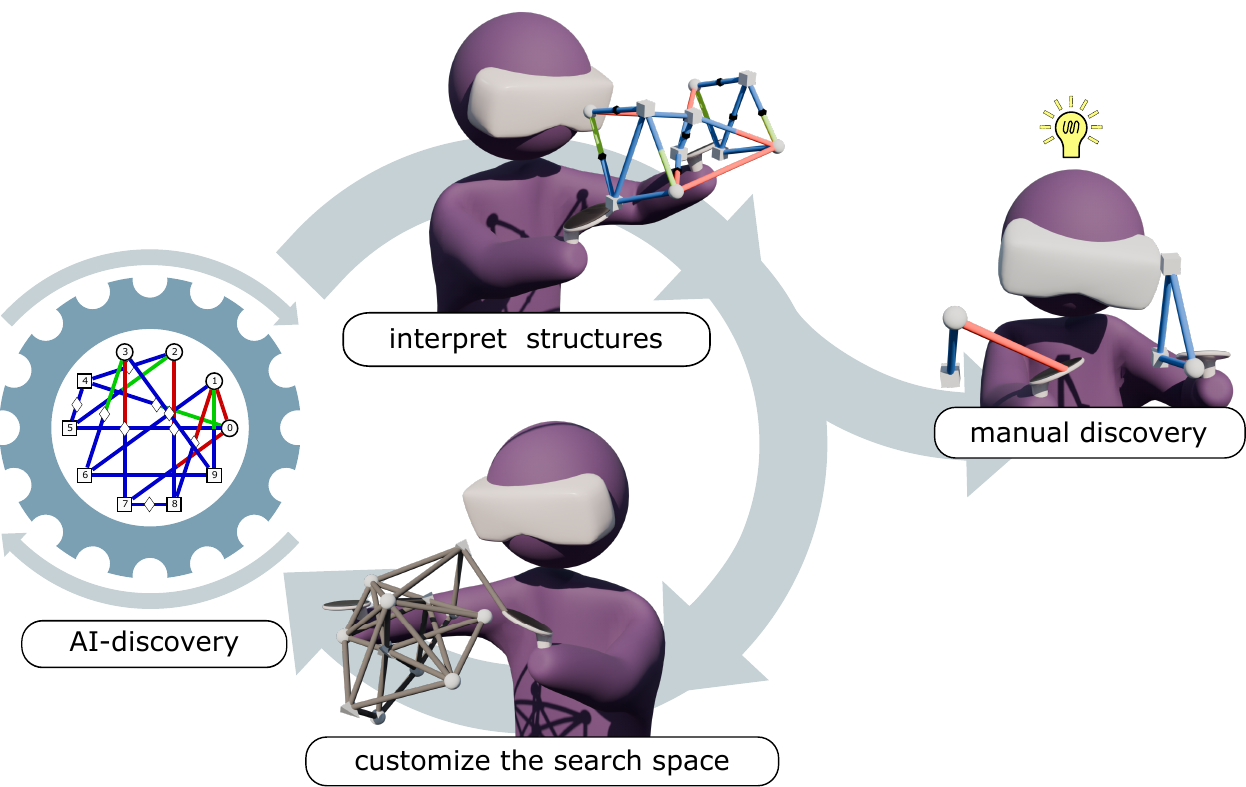}
        \caption{Virtual Reality (VR) assisted digital discovery workflow for obtaining insight from solutions discovered by Artificial Intelligence (AI). AI results are studied in VR utilizing interactive 3D visualization to discover interpretable structures. These are used to manually build new graphs or to direct the AI with smart initial geometries for more efficient searches.}
        \label{fig:VR-workflow}
    \end{figure*}
    
    Virtual Reality (VR) describes the full immersion of a human in a simulated environment, most commonly via the use of a Head Mounted Display, also known as a VR-Headset or VR-Glasses.
    Modern headsets feature stereoscopic rendering on top of the head and gaze tracking combined with gesture controls by tracking controllers in all spatial degrees of freedom.
    The applications of immersive technology are vast and varied \cite{korkut_visualization_2023}, including but not limited to entertainment, education \cite{duer_belle2vr_2018, porter_using_2020, seritan_interachem_2021}, industry and engineering \cite{wolfartsberger_analyzing_2019, qin_application_2020, guo_applications_2020}, neuro- and medical science and therapy \cite{yuan_extended_2023, bohil_virtual_2011, usher_virtual_2018, wang_teravr_2019, steiniger_human_2022, glowacki_group_2022} and data science \cite{donalek_immersive_2014,pirch_vrnetzer_2021}.\\
    
    In the natural sciences, VR is used to visualize and analyze complex, often three-dimensional (3D), scientific data, such as molecular or material data \cite{kutak_state_2023, oconnor_sampling_2018, garcia-hernandez_nomad_2019, kingsley_development_2019, cassidy_proteinvr_2020}, microscopy data \cite{theart_virtual_2017, stefani_confocalvr_2018, spark_vlume_2020, blanc_genuage_2020} or astronomical data \cite{davelaar_observing_2018,sagrista_gaia_2019, baracaglia_e0102-vr_2020}.
    VR is also used to better visualize neural network models \cite{meissler_using_2019, bellgardt_immersive_2020, lyu_aiive_2021} to address the challenge of making Artificial Intelligence (AI) explainable \cite{minh_explainable_2022}.
    In addition, when the AI output is compatible with existing VR frameworks, it can be studied in an immersive environment such as reviewing drug candidates \cite{zhavoronkov_potential_2020}.
    Beyond improved visualization, VR offers natural ways of interacting with scientific results via intuitive gestures \cite{el_beheiry_virtual_2019}, which can be exploited, for example in interactive molecular simulations \cite{oconnor_interactive_2019, lanrezac_wielding_2022}.
    Through this interactivity, the VR platform enables a variety of ingenious methods like using interactive simulations to sample data for machine learning tasks \cite{deeks_free_2023, amabilino_training_2020} or even manipulating microscopic systems in real-time via in-VR-interaction \cite{ferretti_virtual_2021}. \\
    
    In this work, we show that modern VR technology can be effectively applied to enhance the interpretability and conceptualization of scientific results discovered through AI.   
    In particular, we study AI-discovered quantum optics experimental setups.
    By representing such experiments as abstract colored graphs \cite{krenn_quantum_2017, gu_quantum_2019, gu_quantum_2019-1}, we can map complex setups into networks which in turn benefit greatly in legibility and interpretability when visualized n a true 3D environment \cite{pirch_vrnetzer_2021}.
    We demonstrate the unlocked potential by showcasing (1) interpretable structures in AI discoveries and (2) generalizations to new experiments derived from AI results by exploiting their graph geometry.\\
    
    To this end, we built \algo\footnote{In greek mythology Ariadne is Theseus wife, assisting him discovering the way out of the labyrinth. \algo assists the \texttt{Theseus} algorithm \cite{krenn_conceptual_2021} in discovery.}, a webVR tool that allows us to visualize and edit experimental graphs as well as to dynamically compute their current state. \algo is hosted online, and does not require any additional hardware beside of the VR headset and controllers. With it, we discover previously unknown relations in the construction of different state-generation experiments for high-dimensional Greenberger-Horne-Zeilinger (GHZ) states \cite{greenberger_bells_1990} and generalize AI-discovered high-dimensional entanglement swapping experiments to a new setup for high pair counts.
    Furthermore, we extend to human-in-the-loop AI-discovery by intelligently restricting search space based on observed patterns in a related discovery in the case of a high-dimensional GHZ-state analyzer to achieve the ability to measure a larger state.
    The methods used for these discoveries are conceptually depicted in Fig. \ref{fig:VR-workflow}. 
    Our techniques are not limited to quantum optics; they can also be applied to interpret any generative AI outputs involving graph representations.
    
\subsection*{AI-driven Discovery in Quantum Physics}\label{sec:PyTheus_intro}

        \textbf{AI in Natural Science} - 
            In this work, we leverage the potential of VR to augment the process of extracting scientific understanding from solutions to scientific problems discovered by AI.
            Among other tasks, AI is used to efficiently search vast configurational spaces to propose testable answers to research questions \cite{krenn_artificial_2023, wang_scientific_2023}, such as predicting protein folding patterns  \cite{jumper_highly_2021},
            developing drug candidates \cite{arnold_inside_2023}, designing new materials and molecules  \cite{sanchez-lengeling_inverse_2018,celton_deep_2019,nigam_parallel_2022,merchant_scaling_2023}, quantum circuit design \cite{kottmann2023molecular} or discovering experimental setups \cite{krenn_computer-inspired_2020, ruiz-gonzalez_digital_2023}.\\
            
            Even though solutions discovered in this way have value on their own, they might not generally provide scientific understanding if their discovery does not lead to new generalizable insights \cite{de_regt_contextual_2005, regt_understanding_2017, krenn_scientific_2022}.
            Achieving such an understanding can be very challenging when algorithms do not provide conceptual reasoning alongside the output.
            In such cases, the scientist has to deduce the fundamental ideas behind results from examples alone \cite{krenn_scientific_2022}.\\ 
            
        \textbf{Discovery in Quantum Optics} - 
            In quantum optics, algorithms are used to efficiently discover new experiments.
            In its earliest iterations, this is done by assembling a setup out of a pre-defined toolbox of elements to solve a specific task, e.g. designing experiments for quantum state generation, tranformation or metrology \cite{krenn_automated_2016, knott2016search,krenn_computer-inspired_2020}.
            However, many setups can fulfill the same task with minor differences between them, which further expands the already vast search space and decreases search efficiency. 
            This challenge is mitigated by choosing a more efficient, abstract representation of quantum optical setups as colored graphs, where a single graph can represent different setups that yield the same result \cite{krenn_quantum_2017, gu_quantum_2019, gu_quantum_2019-1}.
            Consequently, the search space for experiments is reduced to the space of graphs.
            Additionally, such graphs have been shown to be interpretable, allowing the development of conceptual understanding and novel methods from AI solutions \cite{krenn_conceptual_2021, arlt_digital_2022}.\\
            
        \textbf{The Graph Representation} -
            Here, we briefly explain the graph representation of quantum optics for the specific question of finding experimental setups for generating and measuring photonic quantum systems.
            In this representation, an experimental setup is encoded as a colored, weighted graph by identifying vertices as photon detectors and edges as correlated photon-pair sources \cite{krenn_quantum_2017, gu_quantum_2019, gu_quantum_2019-1}.
            An edge linking two vertices corresponds to the pair-photons propagating to the corresponding detectors.
            The complex edge weight encodes the amplitude and phase of the respective pair. 
            This way, the weights also account for phase shifts and losses in optical
            elements like beam splitters and mirrors, as well as any occurring interference.
            The color of the edge is used to specify the internal photon mode, e.g. its polarization, path, spatial mode, time-bin, or frequency.\\
            
            Experimentally, quantum states are commonly obtained by relying on simultaneous detection events across all detectors, known as n-fold coincidence detection. 
            In the graph picture, such events correspond to subsets of edges containing each vertex exactly once, called 
            \textit{perfect matchings}. As an example, a square graph
                decomposes into two perfect matchings:
            \begin{equation}
                \begin{tikzpicture}[baseline={([yshift=-2pt]current bounding box.center)}]
                \node[shape=circle, draw=black, fill=black, minimum size=1mm, inner sep=0pt] (a) at (0.4,0.4) {} ;
                \node[shape=circle, draw=black, fill=black, minimum size=1mm, inner sep=0pt] (b) at (0.4,-0) {} ;
                \node[shape=circle, draw=black, fill=black, minimum size=1mm, inner sep=0pt] (c) at (0, 0) {} ;
                \node[shape=circle, draw=black, fill=black, minimum size=1mm, inner sep=0pt] (d) at (0,0.4) {} ;

                \path[-](a) edge (b);
                \path[-](b) edge (c);
                \path[-](c) edge (d);
                \path[-](d) edge (a);
                \end{tikzpicture}
            = 
                \begin{tikzpicture}[baseline={([yshift=-2pt]current bounding box.center)}]
                \node[shape=circle, draw=black, fill=black, minimum size=1mm, inner sep=0pt] (a) at (0.4,0.4) {} ;
                \node[shape=circle, draw=black, fill=black, minimum size=1mm, inner sep=0pt] (b) at (0.4,0) {} ;
                \node[shape=circle, draw=black, fill=black, minimum size=1mm, inner sep=0pt] (c) at (0,0) {} ;
                \node[shape=circle, draw=black, fill=black, minimum size=1mm, inner sep=0pt] (d) at (0,0.4) {} ;

                \path[-](a) edge (b);
                \path[-](c) edge (d);
                \end{tikzpicture}
            +                
                \begin{tikzpicture}[baseline={([yshift=-2pt]current bounding box.center)}]
                \node[shape=circle, draw=black, fill=black, minimum size=1mm, inner sep=0pt] (a) at (0.4,0.4) {} ;
                \node[shape=circle, draw=black, fill=black, minimum size=1mm, inner sep=0pt] (b) at (0.4,0) {} ;
                \node[shape=circle, draw=black, fill=black, minimum size=1mm, inner sep=0pt] (c) at (0,0) {} ;
                \node[shape=circle, draw=black, fill=black, minimum size=1mm, inner sep=0pt] (d) at (0,0.4) {} ;

                \path[-](a) edge (d);
                \path[-](c) edge (b);
                \end{tikzpicture}
                \label{eq:pm_decomp}
            \end{equation}
            One such perfect matching corresponds to one term contributing to the final quantum state.
            The amplitude and phase associated with the term are computed as the product of all edge weights in the matching.
            Thereby, a coherent superposition of all the perfect matchings in the graph leads to the final post-selected quantum state.\\
            \begin{figure}[b]
                \centering
                \includegraphics[width=.48\textwidth]{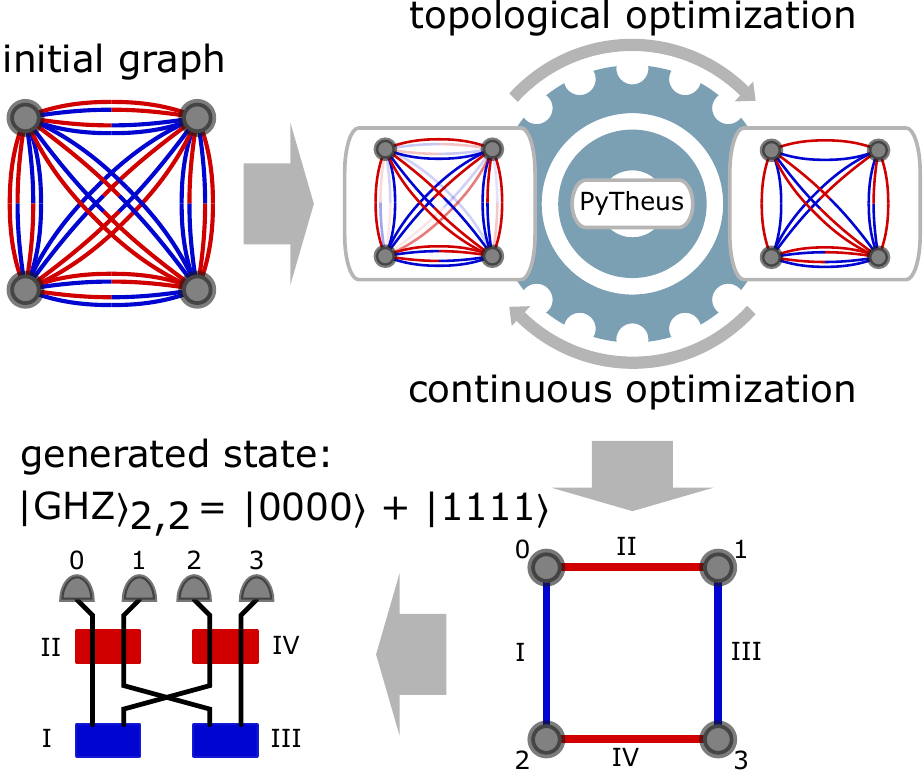}
                \caption{
                    Optimization flow for a 2-dimensional 4-particle Greenberger-Horne-Zeilinger (GHZ) state of the PyTheus-algorithm \cite{ruiz-gonzalez_digital_2023}. Starting from a fully connected initial graph, PyTheus iteratively performs continuous edge weight optimisation and subsequent pruning of unnecessary edges until a minimal graph is reached that fulfills the desired output. 
                    When translating the resulting graph to the experimental setup, every vertex represents a detector, and every edge a correlated photon-pair source emitting towards the connected detectors. 
                    The terms contributing to the final quantum state of the experiment are conditioned on coincidence detection of all detectors. 
                    In the graph picture, these terms correspond to \textit{perfect matchings}, subgraphs where each vertex is only reached by a single edge (see equation \eqref{eq:pm_decomp}).
                    }
            \label{fig:PyTheus_flow}
            \end{figure}
            
            The colored, weighted graph contains the full information of the quantum optical experiment and can be translated into several different realizations using different technologies.
            The most straightforward way of translation is using entanglement by path identity, a way of creating superpositions of emission events by perfectly aligning photon paths \cite{krenn_entanglement_2017, hochrainer_quantum_2022}. Such an example is shown in Fig. \ref{fig:PyTheus_flow}.
            Measurements can be described in the same way using the Choi-Jamiołkowski isomorphism \cite{krenn_conceptual_2021, ruiz-gonzalez_digital_2023}.
            This technique allows for a native translation of experimental setups to abstract graphs and has been practically implemented in recent experimental works \cite{kysela_path_2020,bao_very-large-scale_2023,qian_multiphoton_2023,feng_-chip_2023}.\\
            
        \textbf{PyTheus} -
        PyTheus is an open-source AI tool for discovering new quantum optics experimental setups \cite{ruiz-gonzalez_digital_2023}. It can discover a variety of different types of experiments, such as state generation setups, communication protocols, measurement devices, or photonic quantum gates.\\
        
        The initial step in searching for an experiment is to define the desired experimental task.
        Subsequently, one specifies the initial parameters such as the number of detectors, incoming and outgoing photons, and potential topological constraints.
        With these parameters, PyTheus constructs the largest possible setup in graph form.
        The initial graph is then refined in two iterative steps: (1) by minimizing the desired loss by optimizing the graph weights, and (2) by performing topological optimization to remove unnecessary edges.
        In this way, PyTheus attempts to find the sparsest graph realizing the experiment. 
        The optimization loop is depicted in Fig \ref{fig:PyTheus_flow}.
        
    \subsection*{Physical Background}
        We show graphs of experiments for the creation and measurement of GHZ-states as well as high dimensional entanglement swapping.
        Both topics are highly relevant to the fields of quantum communication and quantum computing but are also used to investigate the fundamental properties of quantum physics \cite{erhard_advances_2020,hu_progress_2023}.
        Here, we briefly introduce the quantum physical background of these experiments.\\
        
        \textbf{GHZ-states and GHZ-analyzers} - 
            GHZ-states are a class of highly entangled multi-particle states originally conceived as a 3-particle generalization of the well-known bell states \cite{greenberger_bells_1990}.
            Both classes of states form a multi-particle orthogonal basis set, where each particle is maximally entangled with all others.
            Since their original proposal, they have been generalized both to particle numbers larger than three and dimensions higher than two \cite{ryu_greenberger_2013,lawremce_rotational_2014,erhard_advances_2020, bao_very-large-scale_2023, xing_preparation_2023}.
            In this work, we discuss graphs relating to the GHZ-state of $n$ particles in $d$ dimensions which is given as:
            \begin{equation}
                |GHZ\rangle_n^d = \frac{1}{\sqrt{d}}\sum_{i=0}^{d-1}\left|j\right\rangle^{\otimes n}.
                \label{eq:nd_GHZ}
            \end{equation}
            GHZ-state analyzers are setups distinguishing one or more GHZ-states out of an unknown input state \cite{pan_greenberger-horne-zeilinger-state_1998}. We discuss binary analyzers distinguishing between the presence or absence of a specific GHZ state given by Equation \eqref{eq:nd_GHZ}.\\
            
        \textbf{High Dimensional Entanglement Swapping} - 
            In its original form, entanglement swapping refers to the preparation of two Bell pairs and the subsequent projection of one particle of each pair in a joined Bell state. This measurement also collapses the remaining two particles in a joined Bell state of their own, without them ever having interacted \cite{zuckowski_event-ready-detectors_1993, pan_experimental_1998}.\\
            This process can be generalized to multipartite entanglement generation using GHZ states as the initial entangled states \cite{bose_multiparticle_1998}. 
            High-dimensional entanglement swapping refers to non-locally creating $d$-dimensional Bell-pairs of the form:
            \begin{equation}
                \left|\Phi^+_d\right\rangle = \frac{1}{\sqrt{d}}\sum_{j=0}^{d-1}\left|jj\right\rangle.
            \end{equation}
            We show graphs creating $n$ such pairs simultaneously between two parties and refer to it as $n$-pair entanglement swapping.\\

Many experiments implementing the tasks mentioned above require additional photons to realize the desired output. These helper photons are called $ancillae$, opposed state photons.
            
\subsection*{AriadneVR}
\label{sec:vr_tool} 
Here we introduce \algo\footnote{\url{https://github.com/artificial-scientist-lab/AriadneVR}}, our web-based virtual reality application for visualizing and analyzing colored graphs representing quantum optics experiments. It is developed with the open-source framework A-Frame\footnote{\url{https://aframe.io/}}  using HTML and JavaScript on an Oculus Quest 2.
The tool runs fully locally in the headset's browser, requiring neither installation nor a tether to a computer for 3D rendering.\\
            
\begin{figure}
                \centering
                \includegraphics[width=\columnwidth]{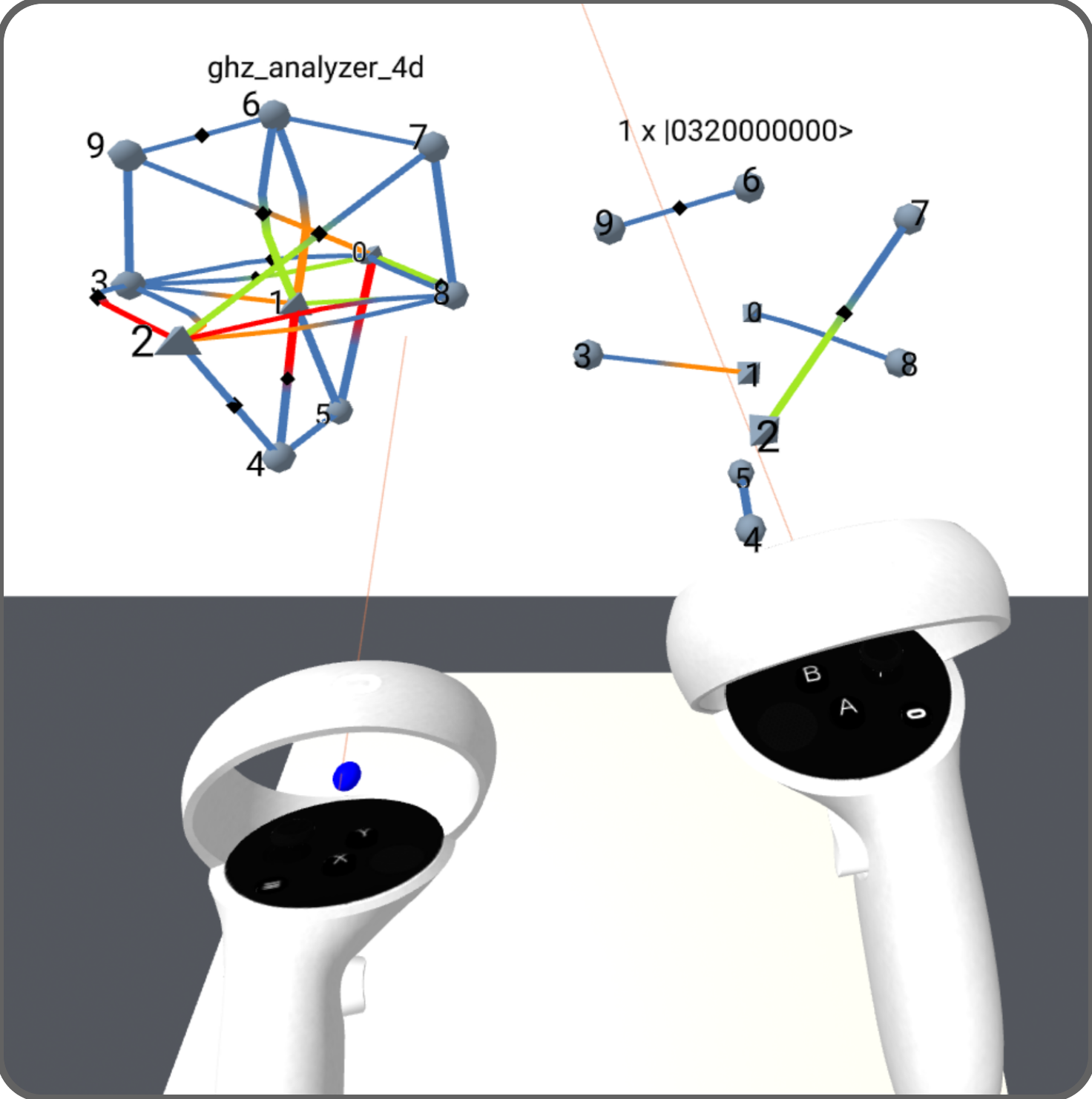}
                \caption{Point of view of an exemplary PyTheus graph analysis inside the environment of \algo. A PyTheus graph is shown with one perfect matching subgraph on the right. A rendered version of this graph is depicted in Fig.\ref{fig:useOf3D}(a). 
                }
                \label{fig:interface}
            \end{figure}

For preprocessing graphs, we use the igraph Python library \cite{csardi_igraph_2005} and its pre-implemented Kamada-Kawai-algorithm \cite{kamada_algorithm_1989} to obtain a 3D-mapping of vertices. This creates an initial arrangement to be perfected manually in virtual reality. 
The layout-generation step is performed outside the tool and creates a file containing all relevant information to be stored on the host platform (e.g. GitHub for hosting via GitHub Pages\footnote{\url{https://pages.github.com/}}).
            When running the tool, A-Frame then uses the THREE.js 3D library\footnote{\url{https://threejs.org/}} under the hood to construct a model from the graph information.
            The VR environment is shown in Fig. \ref{fig:interface}.\\
            
            Our tool allows the exploration of graph representations of quantum optics experiments through visualization and interaction, i.e. manual sorting and editing the experimental setup in its graph form. 
            Beyond this, it supports the steering of PyTheus' searches by generating instruction files based on constrained geometries defined in VR.
            Graphs from the library can be spawned alongside each other and investigated simultaneously.
            The vertex arrangements are changed by simply grabbing and dragging them around as desired.
            When a user makes functional changes to a graph, e.g. adding a new edge by drawing with the controller, the changing state is dynamically tracked.
            The resulting perfect matchings of a graph are available to be spawned as additional 3D models.
            To save the current form of a graph, a file can be downloaded to the headset, which can then be placed into the host library to make it available for later sessions.
            Similarly, the user can download a PyTheus instruction template, where the current graph is set as the initial graph for the search. 
            To streamline this process, edges can be drawn without defining their color, representing a connection where all possible edges are in the initial graph. The framerate for displaying a typical graph is about 40 per second with an Oculus Quest 2.\\
            While \algo is built for analyzing quantum optics experiments, its core functionality is visualizing and editing small colored graphs, hence can be applied to other graphs too, assuming they are fed into the pre-processing pipeline in an appropriate format. As \algo is directly processed at the computer of the VR glass, the framerate will decrease for large graphs. For large graphs, with more than 100 nodes (e.g. graphs studied in social science) a powerful external computer is necessary for rendering the structures. In that case, different software such as VRnetzer \cite{pirch_vrnetzer_2021} are better suited.
        \subsection*{Applications \label{sec:vr_application}}
            \begin{figure}
                    \centering
                    \includegraphics[width=\columnwidth]{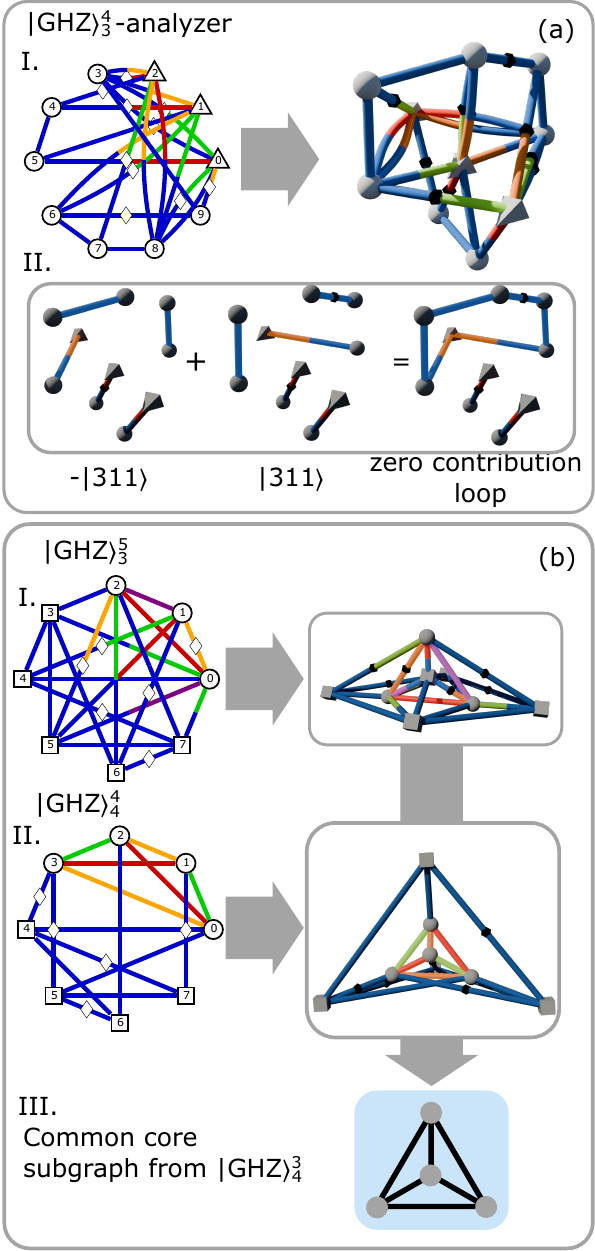}
                    \caption{Through interactive 3D visualization structure is easily revealed in complex graphs. \textbf{(a) I:} 2D representation of a 3-particle 4-dimensional GHZ-state analyzer and its discovered 3D geometry. \textbf{(a) II:} Interference loop example on the undesired $|311\rangle$ ket. \href{https://artificial-scientist-lab.github.io/AriadneVR/4d_3p_GHZ-analyzer}{Interactive Figure}.
                    \textbf{(b) I:} 3-particle 5-dimensional GHZ-state in 2D and 3D. \textbf{(b) II:} 4-particle 4-dimensional GHZ state in 2D and 3D.\href{https://artificial-scientist-lab.github.io/AriadneVR/GHZ-358}{Interactive Figure}.
                    \textbf{(b) III:} Common core subgraph highlighted in both previous graphs signifying the nature of both graphs as generalizations of the $|GHZ\rangle_4^3$ state. \href{https://artificial-scientist-lab.github.io/AriadneVR/GHZ-448}{Interactive Figure.}
                    Ancillae are drawn as squares/cubes, detectors as circles/spheres, and input modes as triangles/tetrahedrons.
                    Colors represent photon modes, indicators on edges their negative weight.}
                    \label{fig:useOf3D}
            \end{figure}
            \begin{figure*}
                    \centering
                    \includegraphics[width=\textwidth]{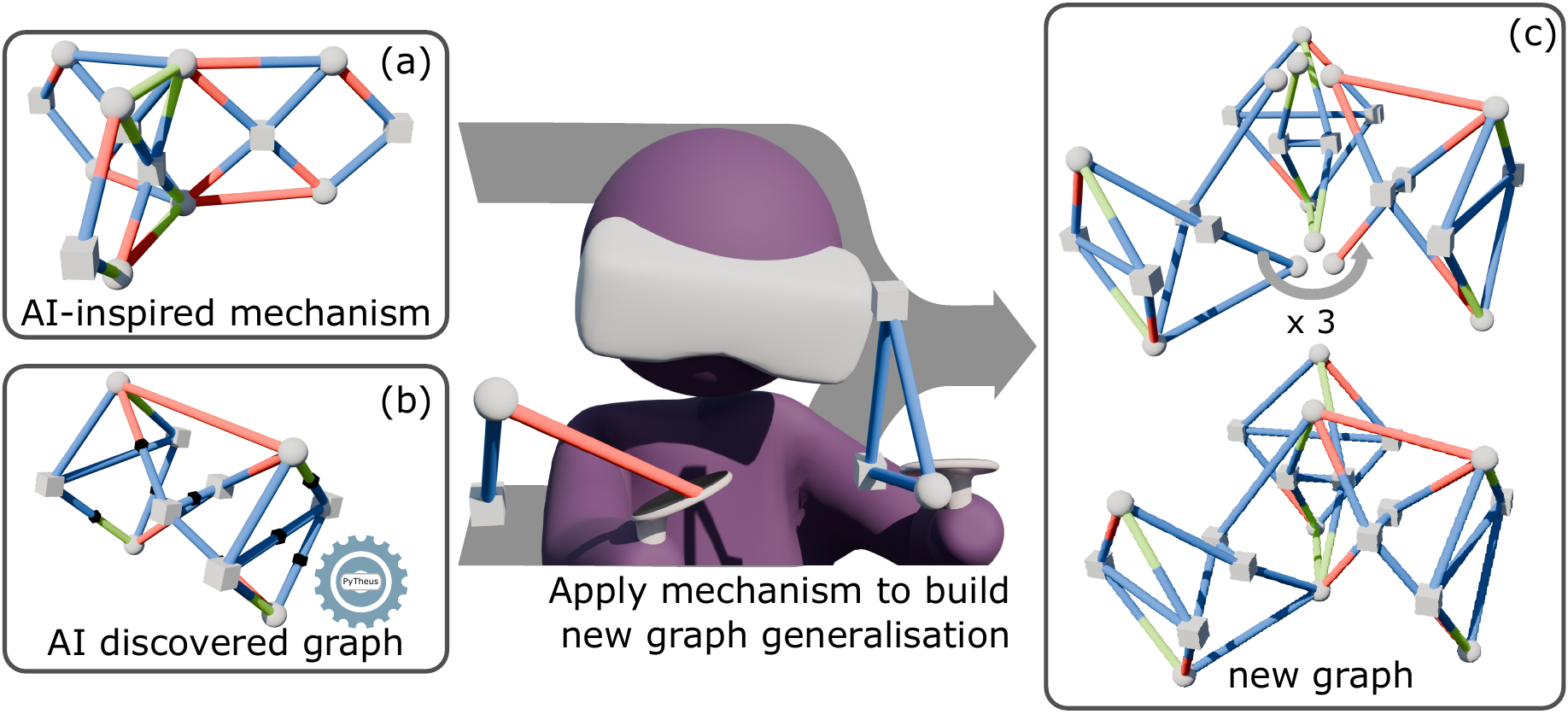}
                    \caption{Manual discovery of new graphs assisted by VR for efficient high dimensional entanglement swapping. PyTheus discovered graphs are watermarked. Through experimentation with graph edits, concepts can be easily transferred and tested on new graphs.\textbf{(a)} Discovered extension mechanism for higher dimensions extracted from efficient PyTheus discovered 2-dimensional entanglement swapping, represented by the graph for three 2-dimensional pairs and a single 3-dimensional pair in the center. \href{https://artificial-scientist-lab.github.io/AriadneVR/trid}{Interactive Figure.}
                    \textbf{(b)} PyTheus discovered solution for 2-pair 3-dimensional swapping.  \href{https://artificial-scientist-lab.github.io/AriadneVR/2p_3d_ES}{Interactive Figure.}\textbf{(c)} Application of the mechanism from (a) to a subgraph from (b) to obtain a new swapping graph for 4-pair 3-dimensional swapping. \href{https://artificial-scientist-lab.github.io/AriadneVR/4p_3d_ES}{Interactive Figure.}}
                    \label{fig:4p3DES_fig}
            \end{figure*}
We use VR in a threefold manner as shown in Fig. \ref{fig:VR-workflow}. The basis of all our VR analysis is the discovery and interpretation of structure in an experimental graph. We provide interactive 3D figures and basic VR environments for all graphs mentioned in this work at the accompanying \href{https://artificial-scientist-lab.github.io/AriadneVR/}{website}. Through manual sorting, the majority of PyTheus graphs can be transformed into visually clean structures, some of which are immediately interpretable, revealing information on how the respective graph achieves its task. If the mechanism underlying a graph is understood, we can manually generalize it to new experiments. This is done directly in VR by drawing new or extending existing graphs. However, in some cases the underlying structure is not immediately obvious, and it remains unclear how to generalize the graph. Here we can use VR to propose geometries of potential generalizations. We then use PyTheus to search for solutions to confirm or reject our candidates.\\
            
\textbf{Finding interpretable structures in 3D Graphs} --
In the following, we show three selected examples from Ref. \cite{ruiz-gonzalez_digital_2023} and Ref. \cite{arlt_digital_2022} where interpretable 3D-representations were found using \algo. For better visual clarity, we use renderings made with Blender 4.0\footnote{\url{https://www.blender.org/}} instead of screenshots of the VR environment for depictions of 3D-graphs.\\    
                
\begin{figure}[h!]
\includegraphics[width = \columnwidth]{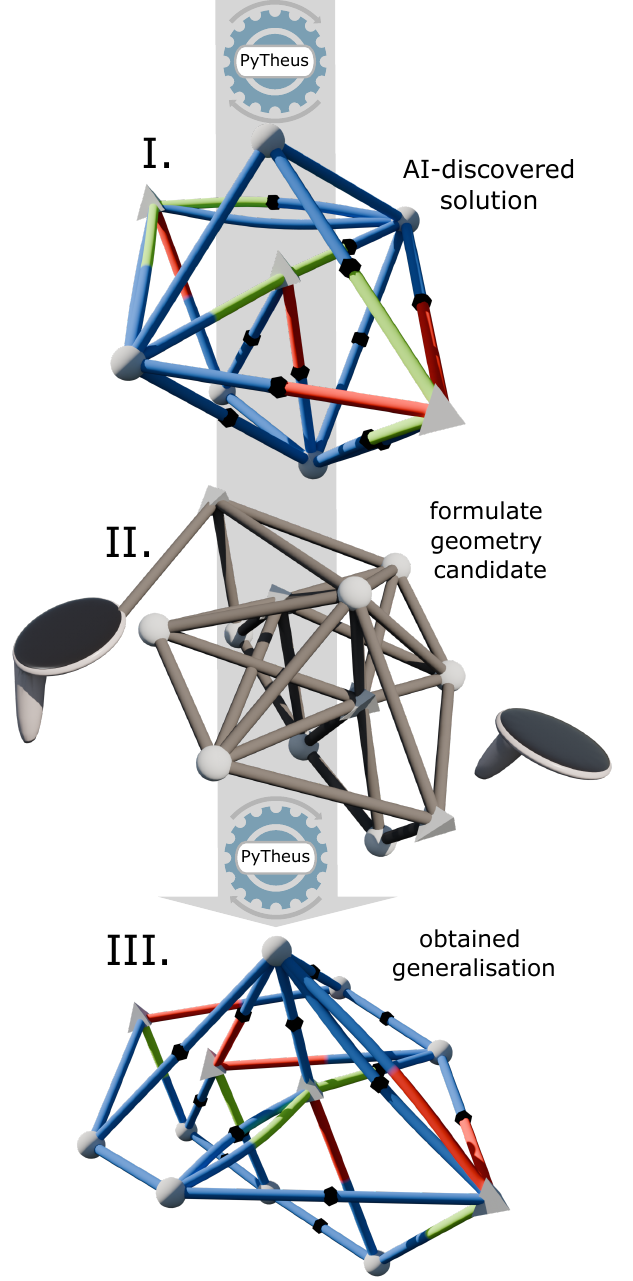}
\caption{Discovering graph generalizations by using \algo to define initial graph geometries for PyTheus searches based on preexisting discoveries. \textbf{I:} Discover the structure in the PyTheus discovered graph of a 3-particle 3-dimensional GHZ-state analyzer. \href{https://artificial-scientist-lab.github.io/AriadneVR/3d_3p_GHZ-analyzer}{Interactive Figure}. \textbf{II:} Use \algo to define a search geometry based on said structure.  \href{https://artificial-scientist-lab.github.io/AriadneVR/3d_4p_GHZ-ana_top}{Interactive Figure}.\textbf{III:} Use PyTheus on the restricted Geometry to discover a 4-particle generalization. \href{https://artificial-scientist-lab.github.io/AriadneVR/3d_4p_GHZ-ana}{Interactive Figure}}
\label{fig:top_extension}
\end{figure}
Fig. \ref{fig:useOf3D}(a) shows a 3-particle 4-dimensional GHZ-state analyzer.
Beyond the striking symmetry of the discovered 3D structure, there is a functional feature exemplified by this highly complex solution.
As a measurement graph, it needs to ensure that all coincidence events produce terms belonging to the desired state or interfere destructively.
            A common way to achieve this in the graph picture is by constructing loops of even edge count with a total negative sign. 
            Such loops were discovered in this particular graph in VR simply by superimposing the 3D models of two interfering perfect matchings. \\
            
            Fig. \ref{fig:useOf3D}(b) shows (I) $|GHZ\rangle_3^5$ and (II) $|GHZ\rangle_4^4$ state creation graphs.
            Ancillae are described by cube-shaped vertices and state photons by spherical vertices.
            Through their 3D structure discovered in VR, it is revealed, that both graphs share the same triangular graph structure highlighted in the center of both graphs. 
            The only difference is the central vertex being an ancilla in $|GHZ\rangle_3^5$ and a state photon in $|GHZ\rangle_4^4$.
            In the latter, this core graph creates a 4-particle 3-dimensional GHZ state.
            The $|GHZ\rangle_3^5$-graph declares one photon of this subgraph as an ancilla (vertex 5 in Fig. \ref{fig:useOf3D}(b) I) to reduce the state photon count from four to three.
            Any remaining photon modes are then added by the outer ancilla structure for both graphs.
            The examples shown here are representative of insights made easily accessible just by moving to stereoscopic rendering and intuitive manual sorting.\\

        \textbf{Manually discovering new AI-inspired experiments in VR} - 
            \algo allows the user to modify and edit graph structures and construct completely new ones.
            Geometric understanding obtained through the superior rendering in VR combined with the user's physical understanding of the experimental mechanisms can then be used to manually engineer new graphs. \\
            
            This process is shown on an example in Fig. \ref{fig:4p3DES_fig}.
            Here, we show the discovery method in the case of resource-efficient high-dimensional multi-pair entanglement swapping.
            The discovery is based on the PyTheus-generated solution for 2-pair 3-dimensional entanglement swapping without requiring a 3-dimensional Bell-state measurement (graph 78 in Ref. \cite{ruiz-gonzalez_digital_2023}) shown in Fig. \ref{fig:4p3DES_fig}(a).
            By combining it with an AI-inspired construction method (see Fig. \ref{fig:4p3DES_fig}(b)) of generating high-dimensional non-locally entangled pairs derived from a different AI-discovered graph for 3-pair 2-dimensional entanglement swapping (graph 77 in Ref. \cite{ruiz-gonzalez_digital_2023}) we can generalize to a new, previously unknown 4-pair 3-dimensional entanglement swapping graph.
            The result is shown in Fig. \ref{fig:4p3DES_fig}(c).
            It effectively creates one whole 3-dimensional pair with no additional resource cost in ancillae by superimposing additional creation processes from single-pair experiments.
            It is an alternative setup to stacking two of the 2-pair AI solutions showing the same novel decrease in required ancillae.
            \algo assists this discovery process by firstly providing easy access to graph structures and secondly offering an intuitive method of defining new graphs through manual drawing.\\
            
        \textbf{Steering AI discoveries through VR via customized search spaces} - 
            Potentially useful generalizations to quantum optical experiments can involve the increase in particle count, dimensionality, or both \cite{erhard_advances_2020, hu_progress_2023}.
            Searching for such generalizations from scratch can be inefficient as the corresponding initial graphs, i.e. the search spaces, quickly increase in size. 
            For example, the initial graph for 4-particle 2-dimensional GHZ-state generation has 24 edges. 
            Searching for a 3-dimensional generalization more than doubles this amount to 54 edges, and attempting a 4-dimensional state without additional ancilla raises the size to 96 edges.\\
            
            We can use \algo to customize our search spaces and decrease the size of the initial graph to steer PyTheus to more efficiently discover desired solutions.
            We achieve this by analyzing existing solutions for lower graph sizes and defining candidate geometries of extensions based on observed patterns.
            In Fig. \ref{fig:top_extension} we show this process for the generalization of a 3-particle 3-dimensional to a 4-particle 3-dimensional GHZ-state analyzer.
            The 3-particle analyzer is highly symmetric.
            Our candidate for a 4-particle geometry keeps this symmetry and follows the same patterns observed in the smaller solution.
            By restricting the initial graph to this geometry, the search for the 4-particle analyzer is reduced from 124 edges down to 74.
            In this way, we can use human intelligence to assist PyTheus in its discovery task, which in turn can provide the human scientist with the example cases needed for concept extraction.

\subsection*{Conclusion and Outlook}
    In this work, we presented ways to leverage the unique potential of immersive technology in the AI-driven discovery process for quantum optics experiments using PyTheus and our purpose-built VR-tool \algo.
    By interactively exploring the graph representations of quantum optics, a new approach to learning from AI to develop new experiments and insight is opened.
    Beyond furthering the understanding of existing solutions for GHZ-state creation and measurement, we achieved substantial machine-inspired discoveries with our tool.
    We engineered a new alternative 4-pair 3-dimensional entanglement swapping setup, where one pair is created at no additional photon cost with knowledge derived from AI discoveries.
    We also showed the human-assisted AI discovery of a 4-particle 3-dimensional GHZ-state analyzer.
    To further exploit the potential of VR our prototype could be expanded to include multi-user support for cooperative analysis and direct access to PyTheus within VR.
    Additionally, a variety of potential visualization features and tools could be added to streamline the analysis process. Examples include computational operations on vertices as well as support for more post-selection conditions.
    Beyond PyTheus' discoveries, the methods developed here can potentially also prove useful in other AI-discovery fields where graph-based representations are used, such as in quantum computing (quantum circuits \cite{kottmann2023molecular}, especially ZX-graphs \cite{duncan_graph-theoretic_2020}), material or protein design (involving crystal structures and molecular graphs \cite{butler_machine_2018}) and DNA-origami design \cite{rothemund_folding_2006, deluca_prediction_2023} in material science and biochemistry.
    We believe that enhancing human researchers' capabilities to comprehend complex data with technologies like VR is crucial to fully utilize the power of AI for scientific discovery. 
    Beyond task-specific applications, immersive technologies also have the potential to fundamentally transform the scientific workspace by allowing remote collaboration and experimental control \cite{gomez-zara_promise_2023}.
\bibliography{AI_discovery,VR-refs,code_refs,VR+AI,phys_bckg}

\end{document}